\documentstyle[aas2pp4]{article}
\input psfig.sty
\begin{document}

\title{ A Numerical Study of Normal Modes of Rotating Neutron Star Models
by the Cowling Approximation}

\author{Shin'ichirou Yoshida and Yoshiharu Eriguchi}
\affil{Department of Earth Science and Astronomy,
	Graduate School of Arts and Sciences,
	University of Tokyo,
	Komaba, Meguro-ku, Tokyo 153-8902, Japan}
\authoremail{yoshida@valis.c.u-tokyo.ac.jp (SY), eriguchi@valis.c.u-tokyo.ac.jp (YE)}

\begin{abstract}
A numerical method of mode analysis of rapidly rotating 
relativistic stellar models by the Cowling approximation 
is applied to rotating neutron stars with realistic 
equations of state. For selected equations of state, eigenvalues 
and eigenfunctions of f-modes are numerically solved for stellar 
models from non-rotating to maximally rotating states.
 
Neutral points of the lower order f-modes are determined as
a function of the stellar rotational frequency. As in the
polytropic case, we find that the bar mode can have neutral
points for models with relatively strong gravity. The rotational
frequency at the neutral point increases as the gravitational mass
of the model becomes larger.

As astrophysical applications of our analysis, we discuss the time scales 
of gravitational radiation induced instability and the possibility 
of the resonant excitation of f-modes during inspiraling 
motion of compact binary systems.

\end{abstract}


\section{Introduction}
In the previous paper (\cite{YE97}, hereafter YE), 
we determined neutral stability points of the
{\it Chandrasekhar-Friedman-Schutz} (CFS) instability of general 
relativistic rotating polytropes, by which non-axisymmetric oscillations
of rotating stars are excited through the coupling with 
gravitational radiation (see \cite{CH70,FS78,JF78}). In the absence of
viscosity this instability sets in at the points where eigenfrequencies 
of the modes vanish as seen from the inertial observer at spatial infinity.
Thus to determine the points on equilibrium sequences of rotating stars
where the model begins to become unstable, zero frequency modes in the 
asymptotically inertial frame must be found.

In YE the equations of state (EOS) of stellar matter were restricted
to the simple relation of polytropes and neutral points of the 
counter-rotating f-modes were obtained.  Here the counter-rotating modes
denote the oscillations whose phase propagation is retrograde with respect 
to the stellar rotation as seen from the observer rotating with the star. 

In this paper we investigate oscillation modes and their neutral points 
of stability for more realistic EOS proposed for the neutron star matter.
The investigated modes are the same ones as in YE which may be 
the most susceptible {\it spheroidal} modes to the CFS instability. 
\footnote{Recent discovery of 
the CFS instability of the r-mode and its strong effect on the stellar 
rotational evolution are also interesting subjects 
(\cite{NA98,AKS98,LOM98}), but they are beyond the scope of this paper.}

It has been discussed that modes with the azimuthal quantum number 
$m=3$ to $5$, by which the eigenfunction of the modes are decomposed
to harmonics having angular $\varphi -$coordinate dependence
$\sim e^{im\varphi}$,
are the most interesting for this instability (\cite{LL86}).
Also interesting is the recently discovered 
bar mode neutral points for rather soft EOS, which never appear in the 
Newtonian framework (for fully general relativistic treatments, see 
\cite{SF97}; also see YE).
Thus we will investigate these lower order modes in this paper.

Concerning the neutral points to the CFS instability, 
Morsink et al. (1998, hereafter MSB) investigated realistic neutron star 
models by applying the numerical method to find the exact neutral modes
of general relativistic rotating stars developed by Stergioulas and 
Friedman (1997). They obtained f-mode neutral points for models with
various masses for several representative EOS. Therefore we will compare 
our results with theirs.

The counter-rotating f-modes are not only interesting in the context of 
the CFS instability of a single neutron star, but also may play an important 
role in compact binary systems because they may couple strongly with the tidal 
potential of the companion. The most significant effect will be the
resonant excitation of the modes by the tidal force and its back reaction 
to the orbital motion of the binary system. We will study this subject
in the last section of this paper.

\section{Brief Summary of the Solving Method}
\subsection{Assumptions}
We assume that axisymmetric equilibrium stars are rotating uniformly
and that the stellar matter is described by zero temperature EOS. 
Under these assumptions, equilibrium states of relativistic rotating 
stars are obtained numerically by the KEH scheme (\cite{KEH89}). 
Linear adiabatic perturbations may be a good approximation
in the present situation, and it is also assumed that the adiabatic index 
$\gamma$ of the perturbation coincides with the local adiabatic index 
of the equilibrium star as follows:
\begin{equation}
	\frac{\epsilon +p}{p}\frac{\Delta p}{\Delta\epsilon}\equiv
	\gamma = \frac{\epsilon +p}{p}
			\left(\frac{dp}{d\epsilon}\right)_{Equil.},
\end{equation}
where $\Delta$ means the Lagrangian perturbation of the corresponding
variable. Eulerian perturbations of the metric components are totally 
omitted as in the previous study (YE), i.e., the Cowling approximation
is adopted.

\subsection{Equations of State}
There exist many candidates for EOS of real neutron stars
with zero temperature.  We here examine some of the representative EOS 
to cover the wide range of stiffness. 
\footnote{See Nozawa et al.~(1998) for recent calculation and summary of 
equilibrium models with various realistic candidates of cold EOS.}
Our choices are those of
1) Pandharipande with hyperon (denoted by EOS B in \cite{AB77}), 
2) Bethe-Johnson without hyperon (\cite{BJ74}), 3) Bethe-Johnson 
(EOS C in Arnett \& Bowers) and 4) more recent WFF3 (\cite{WFF88})
joined to NV (\cite{NV73}) in the low density region.
In order to compare our result with those of MSB, the EOS of Pandharipande 
without hyperon (EOS A in Arnett \& Bowers) is also employed.
Extremely stiff EOS L in Arnett \& Bowers is used only in Figure 8 for the
comparison of the EOSs with a wide range of stiffness.
\subsection{Numerical Treatment}
Our numerical scheme is basically the same as that in YE. Perturbed 
quantities are assumed to behave as $\sim e^{-2\pi i\nu t+im\varphi}$, 
where $t$ is the killing time coordinate. 
A minor change that has been made is the introduction 
of a function $q \equiv \delta p/(\epsilon +p)$, instead of the Eulerian 
perturbation of the Emden function for polytropic stars. Coefficients 
of the perturbed equations contain the background metric and its connection 
coefficients as well as the pressure gradient and a function of adiabatic 
index like $\gamma p/(\epsilon +p)$. \footnote{With these coefficients
introduced, our system of equations is free from coefficients that 
diverge at the stellar surface for relatively stiff EOS unlike 
those of MSB.}

We have used $(r \times \theta) = (100 \times 61)$ grid points for
equilibrium models where $(r, \theta)$ are the spherical polar 
coordinates. Since less number
of grid point is used for the perturbational calculation due to the
restrictions of the power of the computer, values for equilibrium states
are interpolated to give values at the coarse grid points of our 
surface-fitted coordinate (see YE). The interpolation is done by employing
the cubic spline scheme in two dimensions. The results in this paper are 
obtained by using $(r \times \theta) = (25 \times 12)$ grid points in the
surface-fitted coordinate.

\section{Results}
\subsection{Eigenfrequencies and Eigenfunctions}
Rotational sequences of equilibrium stars can be obtained by fixing
the central energy density $\epsilon_c$ and changing the rotational 
parameter such as the ratio of the polar radius to the equatorial 
radius in the meridional cross section of the star. Physical quantities
such as the gravitational mass, the $T/|W|$ value and the angular 
momentum, etc. are calculated after equilibrium configurations are
computed.  Here $T$ and $W$ are the rotational energy and the gravitational
energy, respectively, whose definitions can be found in Komatsu et al. (1989)
and its ratio can be considered to be a standard indicator of the stellar 
rotation.  By changing the central density many sequences 
of equilibrium stars are obtained and a series of models with the same 
gravitational mass and a different rotational frequency can be chosen.
For these equilibrium models the eigenproblem is solved numerically.

In this paper we will concentrate only on the counter-rotating f-modes.
They are the generalization of the Kelvin modes of Newtonian 
Maclaurin spheroids with the counter-rotating phase velocity as seen from
the co-rotating observer with the star and have indices $l=m$ of the 
spheroidal harmonics. These particular modes may be the most susceptible 
to the CFS instability (see \cite{BF86}) and have been mainly investigated 
as to this instability.

Figures 1 -- 7 show the dependency of eigenfrequency $\nu$ on the 
rotational frequency $f$ for the specified EOS and for the fixed 
gravitational mass. For slowly rotating models, the frequency becomes 
higher as $m$ increases. Note that, since we are considering the behavior of 
the perturbed quantities expressed as $\sim e^{-2\pi i\nu t + im\varphi}$, 
the phase velocity $2\pi\nu/m$ is negative in this case.
As the rotational frequency is increased, the rotational dragging effect 
on the mode ({\it not} the 'inertial-frame-dragging' effect in general 
relativity) 
works more strongly for the larger $m$ modes and the order of the mode 
frequencies are reversed. As a result, modes with larger $m$ pass the 
neutral points earlier on the sequences.
\begin{figure}[htbp]
\centering
\psfig{file=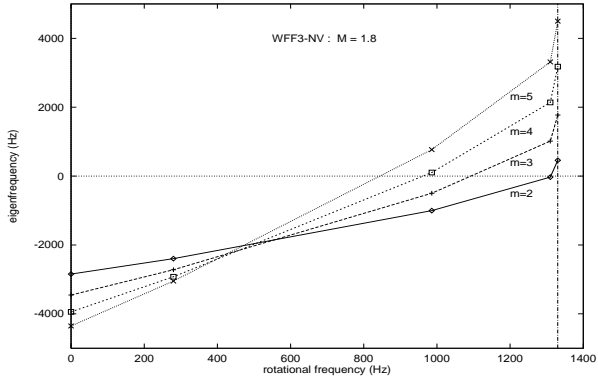,height=5cm,width=8cm,angle=-90}
\caption[fig1.ps]{Eigenfrequencies of f-modes for the neutron stars 
constructed with the WFF3-NV EOS and $M = 1.8 M_\odot$. It should be 
noted that both the eigenfrequency and the rotational frequency 
of the stars are not the angular frequencies but the ordinary frequencies. 
Symbols have the following meanings: '$\Diamond$' for $m=2$ mode, 
'$+$' for $m=3$, '$\Box$' for $m=4$ and '$\times$' for $m=5$. 
The vertical line in the right part is the maximum frequency which
corresponds to the mass-shedding limit of the sequence. \label{fig1}}
\end{figure}

\begin{figure}[htbp]
\centering
\psfig{file=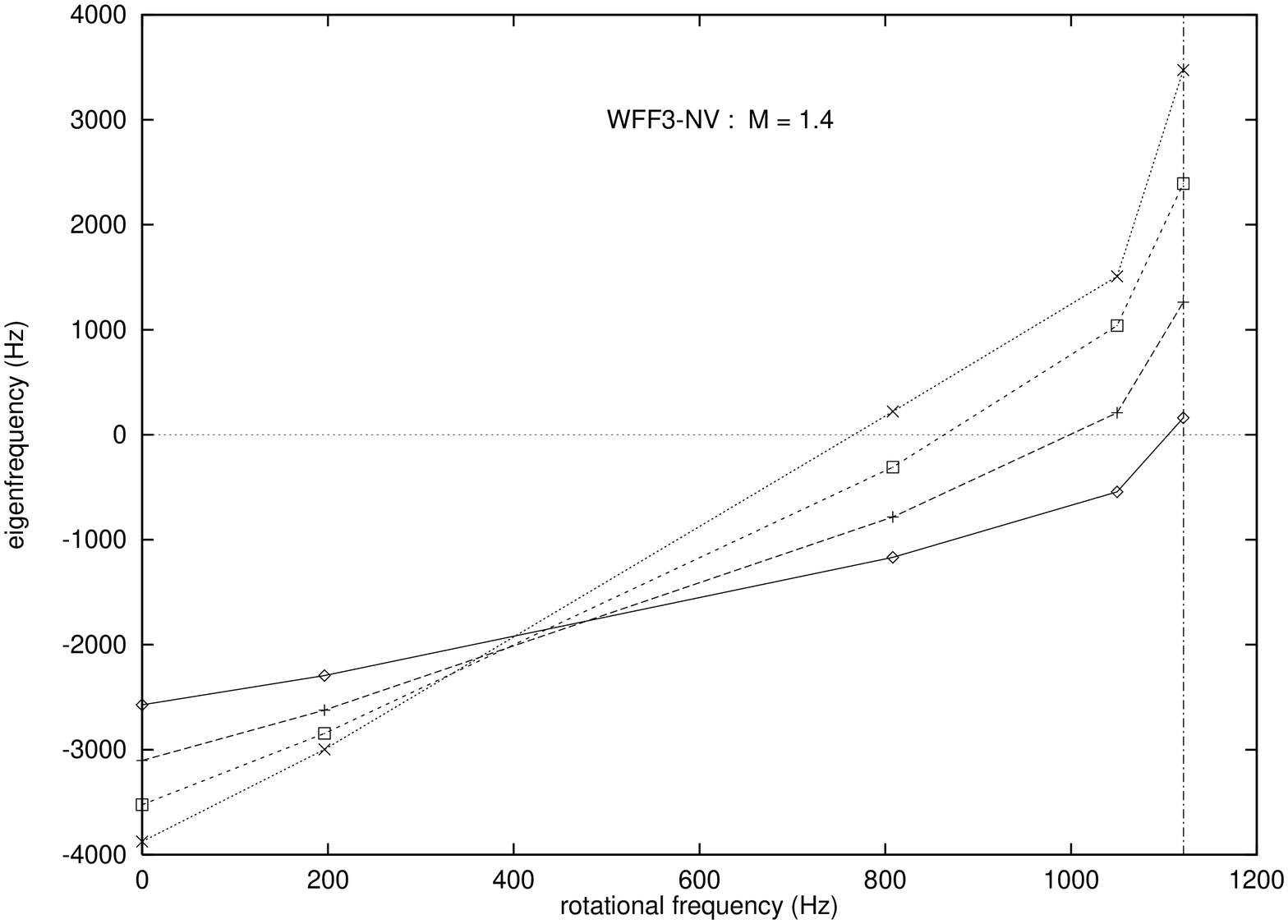,height=5cm,width=8cm,angle=-90}
\caption[fig2.ps]{Same as Figure 1 except for $M=1.4 M_\odot$. \label{fig2}}
\end{figure}

\begin{figure}[htbp]
\centering
\psfig{file=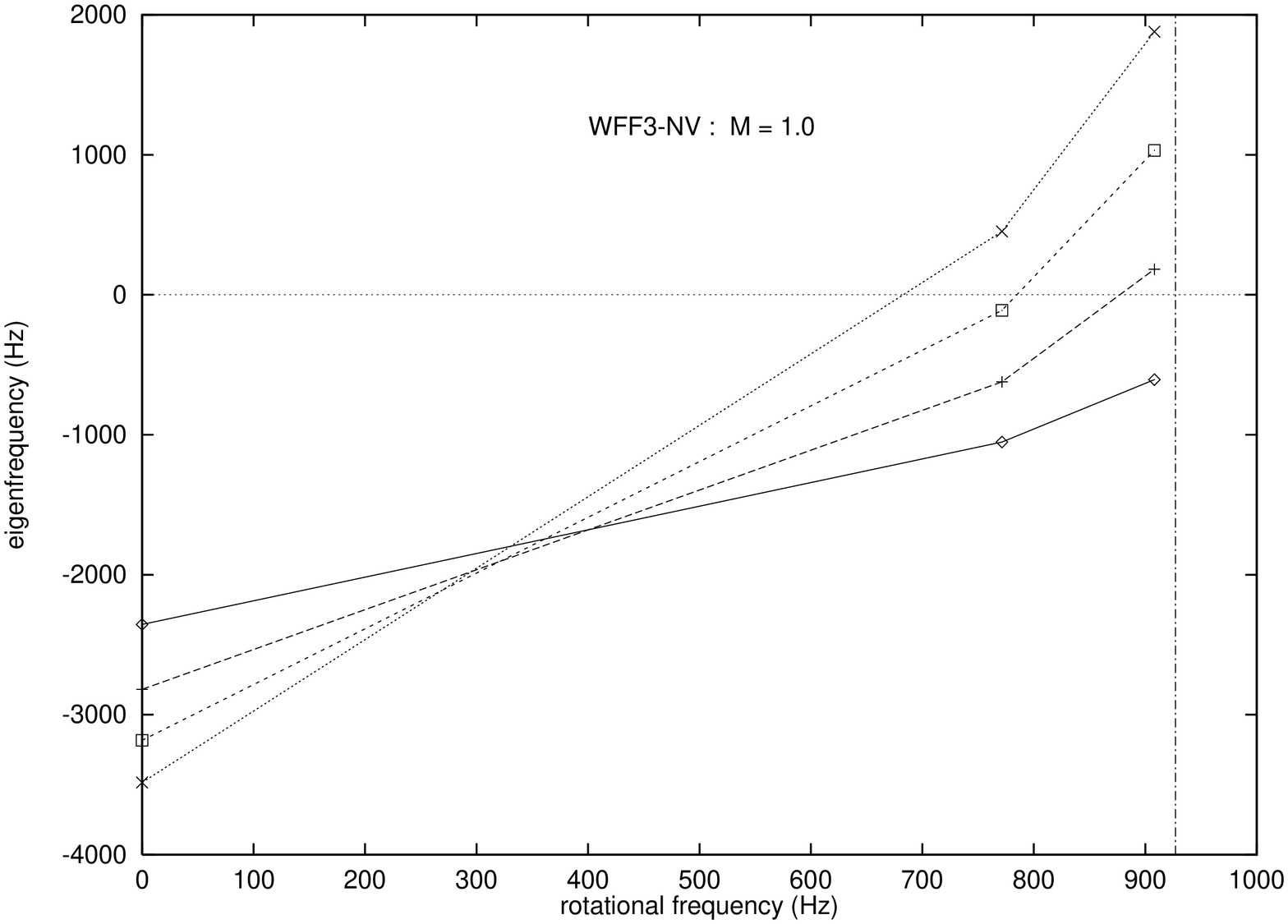,height=5cm,width=8cm,angle=-90}
\caption[fig3.ps]{Same as Figure 1 except for $M=1.0 M_\odot$. \label{fig3}}
\end{figure}

\begin{figure}[htbp]
\centering
\psfig{file=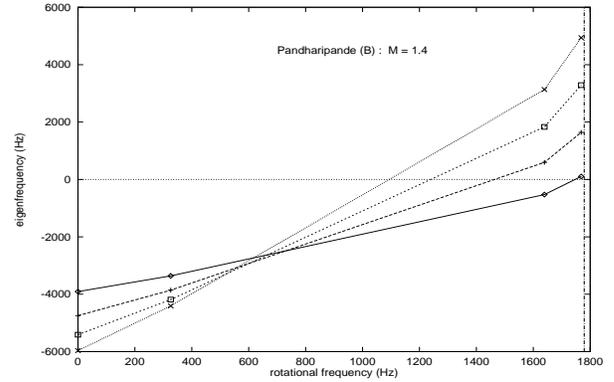,height=5cm,width=8cm,angle=-90}
\caption[fig4.ps]{Same as Figure 1 except for the EOS B and 
$M=1.4 M_\odot$. \label{fig4}}
\end{figure}

\begin{figure}[htbp]
\centering
\psfig{file=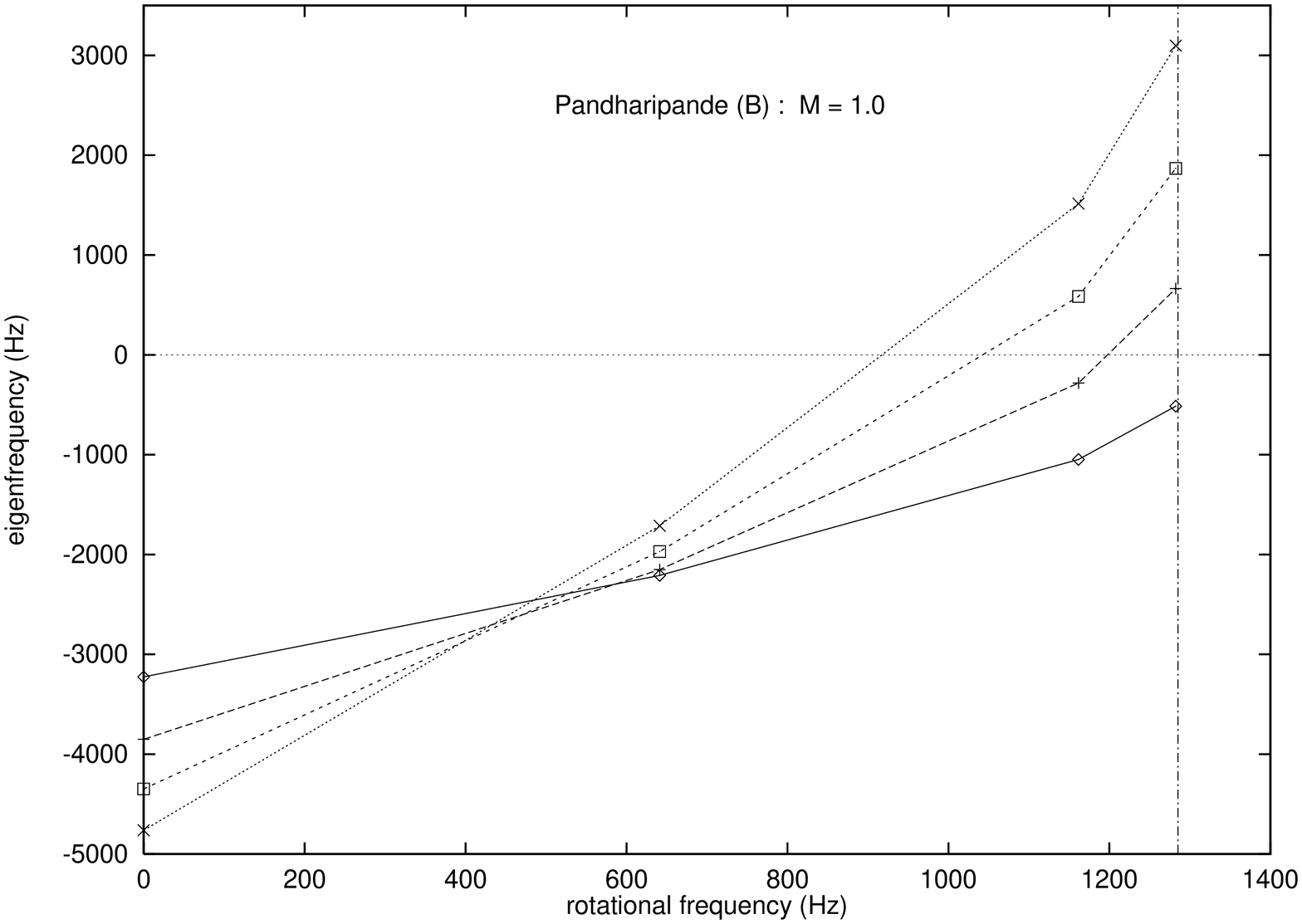,height=5cm,width=8cm,angle=-90}
\caption[fig5.ps]{Same as Figure 4 except for $M=1.0 M_\odot$. \label{fig5}}
\end{figure}

\begin{figure}[htbp]
\centering
\psfig{file=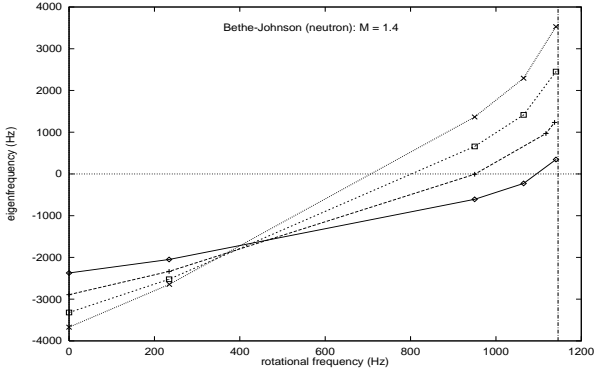,height=5cm,width=8cm,angle=-90}
\caption[fig6.ps]{Same as Figure 4 except for the EOS Bethe-Johnson 
(neutron).\label{fig6}}
\end{figure}

\begin{figure}[htbp]
\centering
\psfig{file=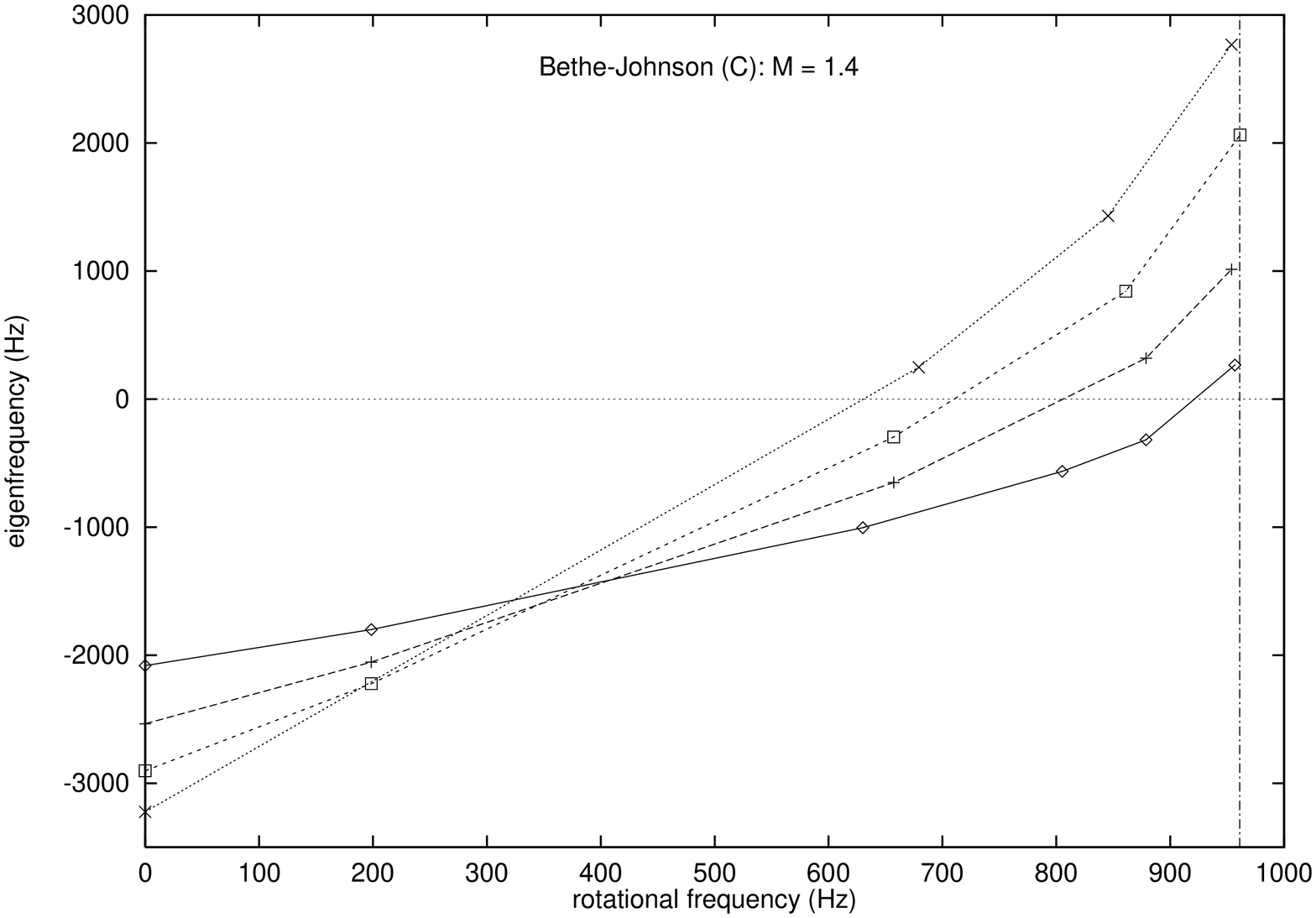,height=5cm,width=8cm,angle=-90}
\caption[fig7.ps]{Same as Figure 4 except for the EOS C.\label{fig7}}
\end{figure}

Near the mass-shedding limit, eigenvalues rise sharply with the increase 
of the rotational parameter. By using more detailed data set from which 
these graphs are produced, we can have much smoother eigenfrequency curves
with the fixed central energy density and show that this behavior is also 
seen there. However, the eigenfunctions of these models show sharp rises 
of the amplitudes near the surface region on the equatorial plane.
Therefore, the rather coarse angular resolution may prevent us from solving 
the numerical eigenvalue problem accurately near the mass-shedding limit.

\begin{figure}[htbp]
\centering
\psfig{file=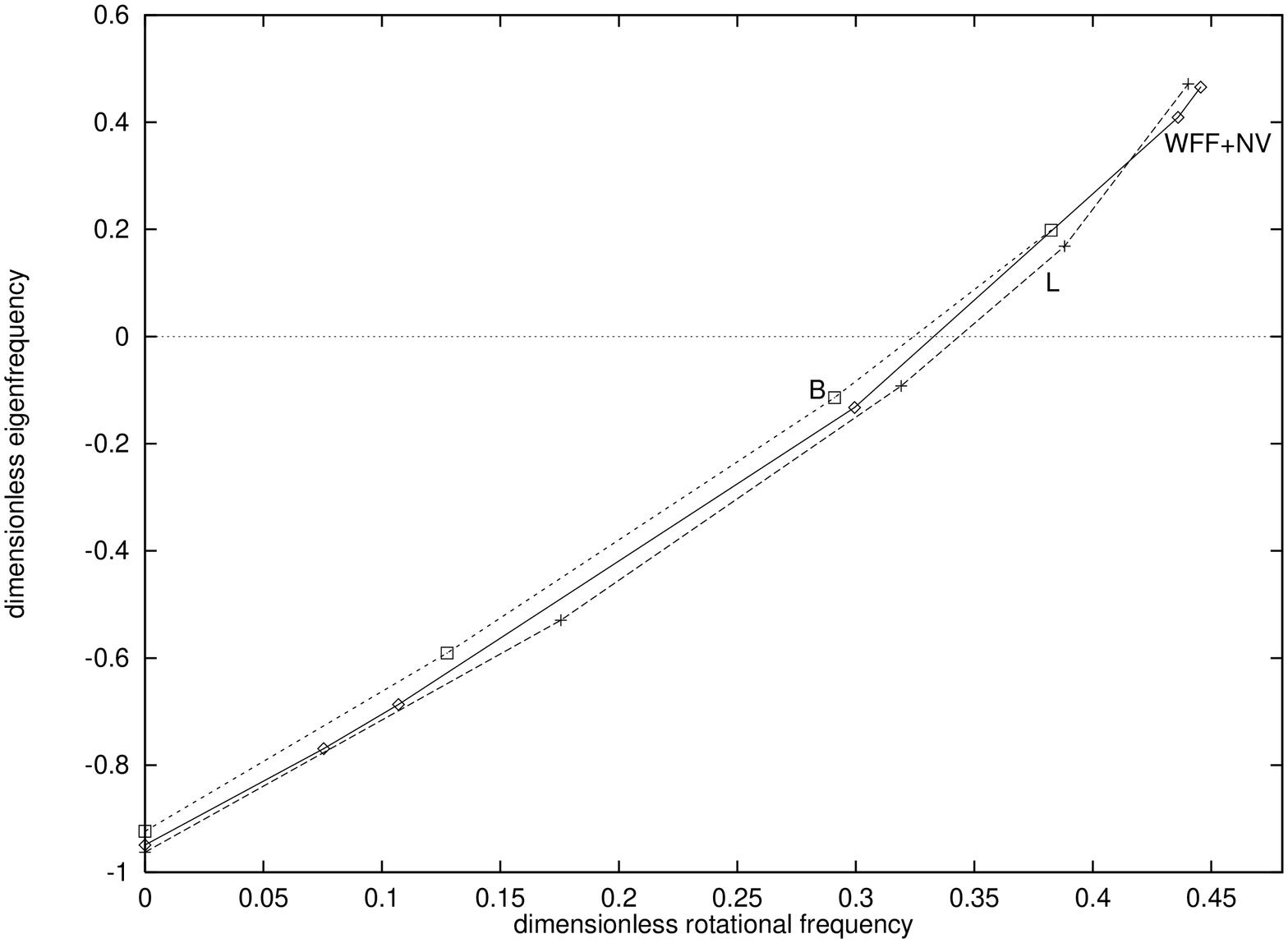,height=5cm,width=8cm,angle=-90}
\caption{Dimensionless eigenfrequency of $m=3$ mode is plotted
against the dimensionless rotational frequency.
The three sequences correspond to the same 
gravitational mass ($M=1.4M_{\odot}$) models with 
different EOSs, WFF3-NV, B and L. \label{fig8}}
\end{figure}

\begin{figure}[htbp]
\centering
\psfig{file=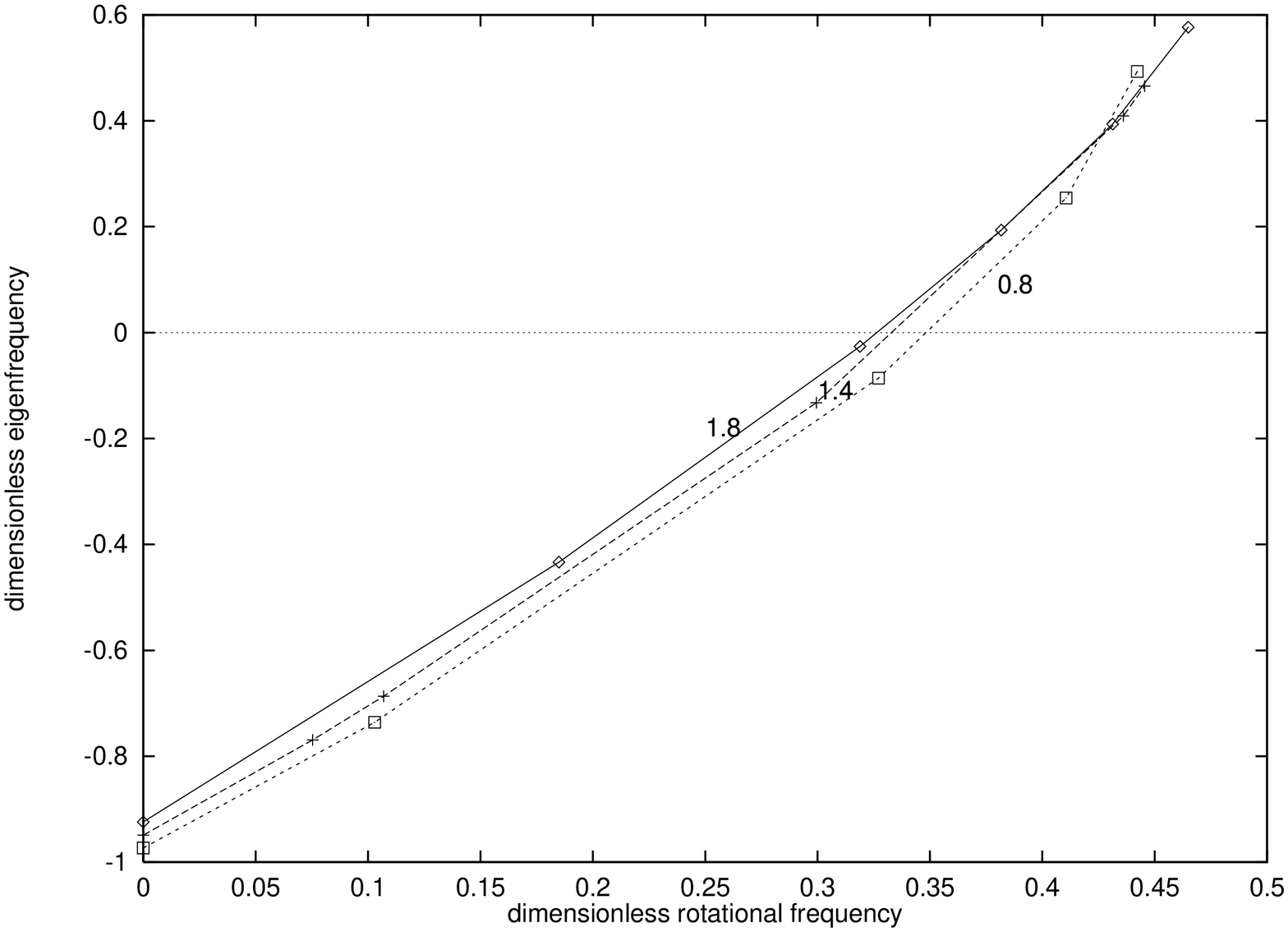,height=5cm,width=8cm,angle=-90}
\caption{Same as Figure 8, except that the EOS is fixed (WFF3-NV)
and the gravitational masses are varied ($M=1.8,1.4,0.8 M_{\odot}$).
\label{fig9.ps}}
\end{figure}

Next we show the typical behaviors of the eigenfrequencies
of the mode of our interest. In Figure 8 dimensionless eigenfrequency
of the $m=3$ mode is plotted against dimensionless rotational frequency 
for three of the EOSs with various stiffness, with the gravitational 
mass of the model being fixed (as $M=1.4M_{\odot}$). 
The EOS B is the softest
among the three, and the EOS WFF3-NV, EOS L become stiffer in this order.
Normalization factor of them is $\sqrt{4\pi\bar{\rho}}$, 
where $\bar{\rho}\equiv M/V_p$ is the averaged density with $M$ and $V_p$
being the gravitational mass and the proper volume of the equilibrium star.
We can see that with this normalization the eigenfrequency is rather 
insensitive to the 
stiffness of the EOS from cases with no rotation to those nearly at the 
mass-shedding limits.

Figure 9 displays the variation of the normalized eigenfrequency due to the 
gravitational mass difference, with the EOS being fixed (as WFF3-NV).
It is seen that increase in gravitational mass makes the mode frequency
larger, which corresponds to the effect of the softening of the EOS in
Figure 8.
This is reasonable since the strong self-gravity of the star 
tends to induce its density profile to be concentrated to the 
central region of it, which effectively realizes the configuration 
with the softer EOS.

In Figures 10 and 11, we show the typical behavior of the eigenfunction 
$q$. The equilibrium models compared in these two figures have the same 
central density but the rotational frequencies are different. 
In both models, the function $q$ increases monotonically 
from the center to the surface of the star along the radial spokes 
in the surface-fitted coordinate. For the slowly rotating model (Fig.10), 
the angular dependence of the $q$ is nearly that of the associated 
Legendre function, $P_l^m(\cos\theta)$ (in this case $l=m=4$), 
whereas rapid rotation tends to shift the distribution of the 
function $q$ to the equatorial plane (Fig.11). 
\begin{figure}[htbp]
\centering
\psfig{file=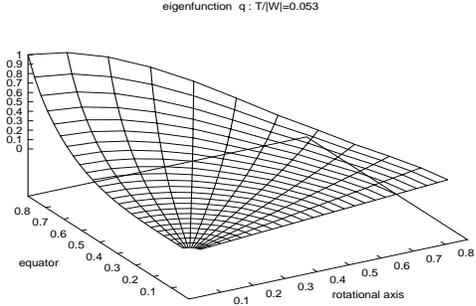,height=5cm,width=8cm,angle=-90}
\caption{An example of the eigenfunction $q$ (see text) for 
a slowly rotating star with the WFF3-NV EOS. A quarter of the meridional 
cross section of the star is shown. The mode number $m=4$. 
The radial coordinate distance is normalized by using 
$c/(4\pi\epsilon_c)^{1/2}$ where $c$ is the velocity of light 
and $\epsilon_c$ is the central mass density of the star. 
The amplitude of the eigenfunction is normalized such that the value of 
$q$ at the surface point in the equatorial plane becomes unity. 
The parameters of the equilibrium model are: 
$\epsilon_c=1.0\times 10^{15}$g/cm$^3$, 
the rotational frequency $f=285$ Hz,
$M=1.19 M_\odot$ and $T/|W|=0.053$. The eigenfrequency $\nu =-2391$ Hz. 
\label{fig10}}
\end{figure}

\begin{figure}[htbp]
\centering
\psfig{file=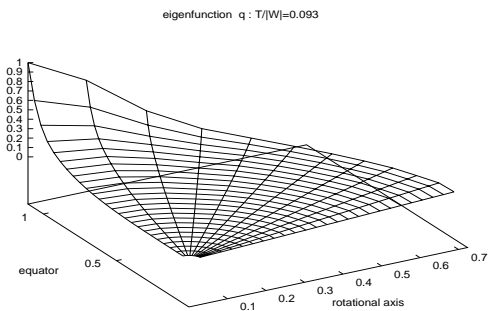,height=5cm,width=8cm,angle=-90}
\caption{The eigenfunction $q$ for a rapidly rotating star.
The same mode as in Figure 10.
The parameters of the equilibrium model are:  
$\epsilon_c=1.0\times 10^{15}$g/cm$^3$,
$f=1092$ Hz, $M=1.43 M_\odot$ and $T/|W|=0.093$. 
The eigenfrequency $\nu =1250$ Hz. \label{fig11}}
\end{figure}

To see the 'radial' dependence of the eigenfunction, we show the 
function $q$ on the equatorial plane (Fig.12). Here the same models
are used as those in Figures 10 and 11. As a 'radial' coordinate here we 
take the value $x\equiv 1-\epsilon/\epsilon_c$, which roughly represents 
the matter-energy distribution of the equilibrium stars.
As seen in this figure, the distribution of the function $q$ is concentrated
to the surface region as the star rotates more rapidly. 
In the Newtonian theory, the same situation seems to improve the 
Cowling approximation for rapidly rotating stars (\cite{SY97}) 
because the role of perturbed gravitational potential becomes less
important. Roughly speaking, less part of the stellar mass participates 
in the oscillation as the star rotates more rapidly. This is also 
likely to be the case in general relativistic stars, though we have no 
'exact' quasi-normal modes to compare with.
\begin{figure}[htbp]
\centering
\psfig{file=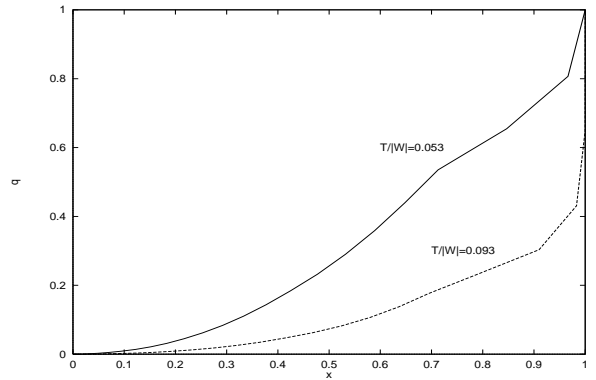,height=5cm,width=8cm,angle=-90}
\caption{The eigenfunction $q$ on the equatorial plane 
is plotted against the density coordinate $x$. The equilibrium models
are the same as in Figure 10 and 11. Mode number $m=4$.
The solid curve is that for the model with $T/|W|=0.053$, whereas the 
dashed one is that for the model with $T/|W|=0.093$. \label{fig12}}
\end{figure}

In Figure 13 the differences of radial behavior in the equatorial 
plane between modes with different $m$ are shown. Here the radial 
coordinate is that of the surface-fitted coordinates. It can be seen 
that as $m$ increases, the main part of oscillation of the star shifts 
outward where the amount of the mass fraction decreases. This can 
make the Cowling approximation more accurate for higher order modes 
(see Table 1).
\begin{figure}[htbp]
\centering
\psfig{file=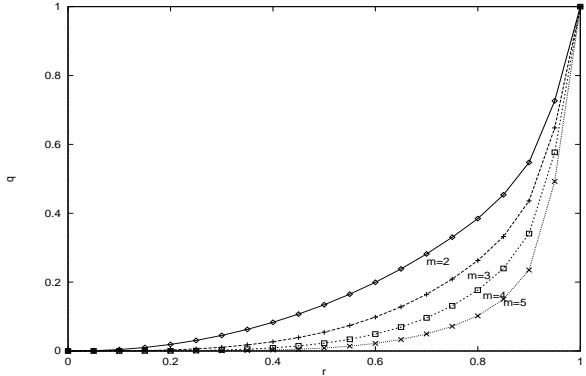,height=5cm,width=8cm,angle=-90}
\caption{The eigenfunction $q$ on the equatorial plane with 
different mode numbers. The radial coordinate distance is that of the 
surface-fitted coordinate. The equilibrium model is the same as 
that in Figure 11. Symbols have the same meanings as in Figure 1. 
\label{fig13}}
\end{figure}
%
\subsection{Neutral Points of the CFS Instability}
As is already remarked, we can estimate the neutral points of the
CFS instability by finding zeroes of the 
eigenfrequencies. \footnote{Note that the neutral points here are 
{\it not} determined by using the graphs shown in the previous 
section. More detailed data sets have been used to obtain them.}
In Table 1 we compare the values of $T/|W|$ at the neutral points 
of the instability with those obtained by MSB. 
%
%
Here we summarize 
the tendency of the Cowling approximation in general relativity as
follows:
(1) for the same EOS and for the same mode number, the Cowling approximation
    gives better results as the central energy density of the model
    increases; 
(2) for the same equilibrium model, it results in more accurate values
    for larger mode numbers; and 
(3) the Cowling approximation has a tendency to overestimate
	the stability in the case of relatively weak gravity 
	(cf. the Newtonian case in \cite{SY97}).
These are qualitatively consistent with the previous results in YE. 
In contrast to the property (3), for larger central density 
(strong gravity) models, higher order modes seem to be underestimated 
in its stability by the Cowling approximation. As for the bar mode 
there seems no improvement with increase of the central density. 

\begin{table}[hbt]
\caption{Comparison of values of $T/|W|$ at neutral points with 
the results by Morsink et al. (1998) \label{tabl1}}
\vspace{0.2cm}
\begin{center}
\begin{tabular}{ccccc}\tableline
EOS&mode&$\epsilon_c$($\times 10^{15}$g/cm$^3$)& present & MSB\\\tableline
    A & $m=2$ & 1.0 & 0.094 & 0.082\\
      &       & 3.2 & 0.079 & 0.056\\
      & $m=3$ & 1.0 & 0.081 & 0.066\\
      &       & 3.2 & 0.049 & 0.044\\
      & $m=4$ & 1.0 & 0.056 & 0.054\\
      &	      & 3.2 & 0.035 & 0.035\\
      &	$m=5$ & 1.0 & 0.043 & 0.044\\
      &	      & 3.2 & 0.027 & 0.029\\ 
&&&&\\
    C & $m=2$ & 0.74 & $-$ & 0.087\\
      &	      & 0.90 & 0.098 & 0.082\\
      &       & 2.5 & 0.077 & 0.059\\
      &	$m=3$ & 0.70 & 0.076 & 0.066\\
      &	      & 0.95 & 0.071 & 0.061\\
      &	      &	2.5 & 0.048 & 0.046\\
      &	$m=4$ & 0.70 & 0.053 & 0.052\\
      &	      &	1.0 & 0.049 & 0.047\\
      &	      & 2.5 & 0.035 & 0.036\\
      & $m=5$ & 1.0 & 0.036 & 0.038\\
      &	      & 2.5 & 0.027 & 0.028\\\tableline

\end{tabular}
\end{center}
\end{table}

When the EOS is fixed, we can calculate an eigenfrequency of a mode
for the model with a given rotational frequency $f$ (Hz) and 
a given gravitational mass $M$ ($M_\odot$). 
Then we have a neutral stability curve of 
the CFS instability for the mode in the $f - M$ plane as a set of 
zeroes of the eigenfrequencies.
Four neutral stability curves corresponding to four different modes
with different EOS are shown in Figures 14 -- 17.
\begin{figure}[htbp]
\centering
\psfig{file=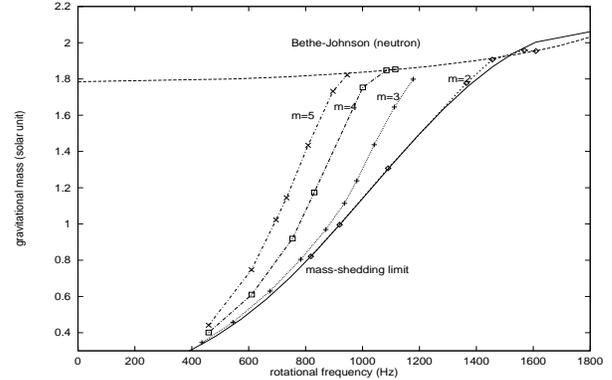,height=5cm,width=8cm,angle=-90}
\caption{Neutral stability curves of the f-mode in the 
$f - M$ plane. The gravitational mass is normalized by the solar mass. 
The EOS is that of Bethe-Johnson without hyperon contribution.
The solid line is the mass-shedding limit curve. 
The dashed line is the approximate maximum mass curve for a given
rotational frequency (see text). Symbols used here are the same as in 
Figure 1. \label{fig14}}
\end{figure}

\begin{figure}[htbp]
\centering
\psfig{file=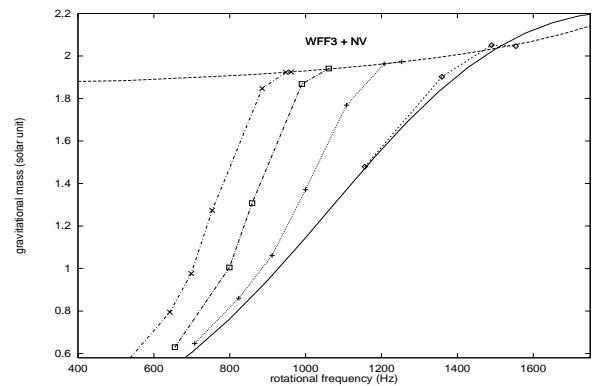,height=5cm,width=8cm,angle=-90}
\caption{Same as Figure 14 except for the WFF3-NV EOS. 
\label{fig15}}
\end{figure}

\begin{figure}[htbp]
\centering
\psfig{file=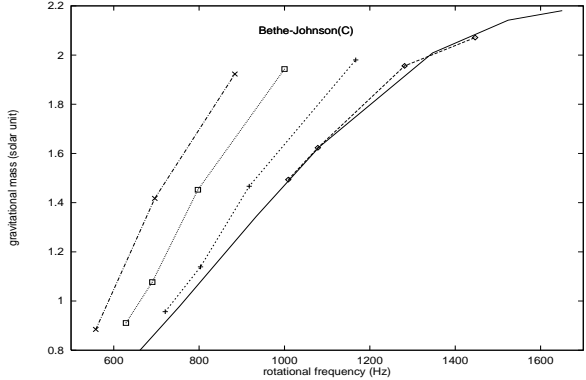,height=5cm,width=8cm,angle=-90}
\caption{Same as Figure 14 except for the EOS C. 
The approximate maximum mass curve is omitted here. \label{fig16}}
\end{figure}

\begin{figure}[htbp]
\centering
\psfig{file=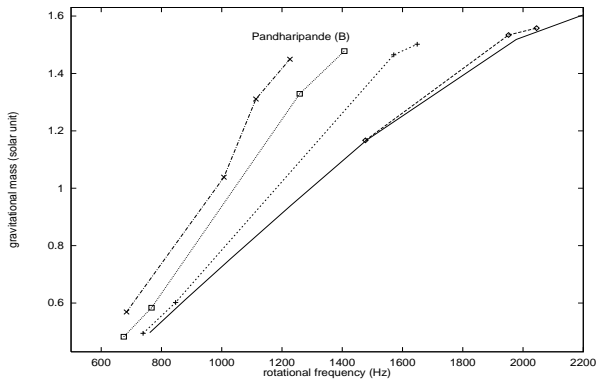,height=5cm,width=8cm,angle=-90}
\caption{Same as Figure 14 except for the EOS B. \label{fig17}}
\end{figure}

Figure 14 displays those for the EOS of Bethe-Johnson without hyperon.
The region of the left hand side of each curve is the stable region
against the CFS mechanism for the corresponding mode. 
The solid line is the mass-shedding limit curve
on which the gas on the surface in the equatorial plane rotates with the local
Keplerian velocity. The dashed line in the upper region is the approximate 
maximum mass line for a given rotational frequency. The reason why we use 
the word 'approximate' is that to obtain it we do not constrain the 
rotational frequency to be constant but the axis ratio of the configuration 
to be constant (\cite{KEH89}). 

At first sight, it seems strange that for sufficiently rapidly rotating 
cases we have a model in the right hand side region of the mass-shedding 
curve. It seems to imply that for a given mass we have equilibrium models 
that have larger rotational frequencies than the mass-shedding case.
This comes from the fact that the rotational frequency $f$ is {\it not}
a proper measure of the rotation for extremely rapidly rotating
stars (see appendix for its explanation).

\section{Applications}
We here apply our results to two issues of interest in neutron star
physics. One is the estimation of the time scale of the 
CFS instability. The other is related to the resonant excitation of 
f-modes in inspiraling compact binary systems which are possible targets 
of the gravitational wave detectors under construction.

\subsection{CFS Instability of Neutron Stars}
In the previous section neutral points of the CFS instability are 
determined. Then what we want to know next is {\it how fast these unstable
perturbations grow beyond these points}. Unfortunately our approximation
does not provide us the answer directly.
What is expected to happen in the real process is that the stellar
free oscillations with non-zero frequencies couple to gravitational wave
radiation which carries its energy to infinity, and the frequencies inevitably 
have imaginary parts. This is the problem of the so-called 
{\it quasi-normal mode} which is well-known in black hole physics.
To obtain complex eigenfrequencies, we must take the metric
perturbations into account and the {\it out-going wave} conditions must be
specified to them. This task is extraordinarily difficult for rapidly
rotating stars and no investigation has been accomplished for it yet.

As its alternative, there have been several works which estimates the 
growth time of the instability by applying the gravitational radiation 
reaction potential of the post-Newtonian expansion (\cite{KT69}) to 
Newtonian stellar oscillations. In this approximation, we only need to 
know the time dependency of the mass multipole of the oscillating star. 

Here we roughly estimate the growth rates of the instability by following
Comins (1979), who investigated the secular effects of the gravitational
radiation reaction and viscosity on the oscillating Maclaurin spheroids.

According to the analysis of Comins, the effect of gravitational radiation 
adds a small imaginary part to the oscillation frequency viewed from the 
co-rotating frame with the star $\Sigma\equiv -2\pi (\nu-f)$ as follows:
\begin{equation}
 \delta\Sigma = \frac{2iG\left[\frac{M}{\frac{4}{3}\pi R_1^3}\right]
			(m+1)(m+2)R_1^{2m+1} (2\pi)^{2m}
			\nu^{2m+1}}
			{(m-1)\left[(2m+1)!!\right]^2
			\left[(m-1)f - \nu\right]c^{2m+1}},
\end{equation}
where $c$ and $G$ are the velocity of light and the gravitational constant,
respectively, and $R_1$ is the equatorial radius of the star.
From this expression, we can estimate the e-folding time $\tau_{GR}$
for the growth of perturbations as follows:
%
\begin{eqnarray}
 \tau_{GR} (\mbox{sec}) &=& k(m)
	\left[ (m-1)\left( \frac{f}{\mbox{kHz}} \right) 
		- \left(\frac{\nu}{\mbox{kHz}}\right)\right]\times\nonumber\\
&&	\left(\frac{M}{M_{\odot}}\right)^{-1}
	\left(\frac{R_1}{10\mbox{km}}\right)^{-2m+2}
	\left(\frac{\nu}{\mbox{kHz}}\right)^{-2m-1},
\end{eqnarray}
%
where $k(m)$ is defined as,
\begin{equation}
k(m) = 	\frac{10^{2m-4}\left[c/10^{10}\right]^{2m+1}
	(m-1)\left[(2m+1)!!\right]^2}
	{4(2\pi)^{2m-1}(m+1)(m+2)}.
\end{equation}

This formula is applied to the neutron star models of 
$M=1.4M_{\odot}$ with the WFF3-NV EOS (Table 2). 
%
%

It is seen
that, as the rotation rate is increased, the lower order modes suffer 
stronger destabilization effect by gravitational radiation, which 
originates from the efficiency of gravitational radiation of these lower 
order oscillations, than the higher order modes whose
neutral points reside at the lower rotational frequencies.
As a result these modes shown here have the same order of 
timescale near the mass-shedding limit.

\begin{table}[hbt]
\caption{Estimated timescale in units of second 
of the CFS instability for the $M=1.4 M_{\odot}$
star with the WFF3-NV EOS. \label{tabl2}}
\vspace{0.2cm}
\begin{center}
\begin{tabular}{ccccc}\\\tableline
$f$(Hz) && $m=3$ & $m=4$ & $m=5$\\\tableline
$818$ && $-$ & $-$ & $9\times 10^{15}$\\
$952$ && $-$ & $-$ & $2\times 10^{9}$\\
$1006$ && $4\times 10^{19}$ & $4\times 10^7$ & $6\times 10^7$\\
$1053$ && $3\times 10^7$ & $1\times 10^6$ & $3\times 10^6$\\
$1117$ && $1\times 10^3$ & $1\times 10^3$ & $3\times 10^3$\\\tableline
\end{tabular}
\end{center}
\end{table}

This estimation needs to be treated as very rough one because we assume 
that the formulae for Newtonian Maclaurin spheroids are applicable to the
relativistic neutron star models and because the correspondence of equilibrium
quantities such as the mass and the equatorial radius is vague.
However, qualitative behavior of the modes would be the same
even if more refined treatment would be employed.

\subsection{Resonant Excitation of the f-modes for Inspiraling Compact 
Binary Systems}

Inspiraling compact binary systems (neutron star--neutron star (NS/NS),
black hole--neutron star (BH/NS) and black hole--black hole)
are the most promising sources of gravitational wave for large 
interferometry gravitational wave detectors under construction 
such as LIGO and VIRGO (see e.g. \cite{KT94} and the references therein). 
These detectors will be able to
observe the inspiraling phase where components of the system are well 
approximated by the point masses. To extract meaningful results from the 
gravitational wave signals, it is indispensable to have sufficiently 
accurate theoretical templates of wave forms in the frequency range 
of $10-10^3$Hz to which these detectors are sensitive (\cite{CET93}).

In this context, if at least one of the components is a neutron star, 
its internal hydrodynamical degrees of freedom may be a potential threat 
to the template construction. Bildsten \& Cutler (1992) examined
the problem of the 'equilibrium tide' and the issue of the tidal locking of the
components. The tidal locking which causes synchronization of the spin
and the orbital motion seems rather unlikely to occur according to their 
result, and the theoretical template suffers only a negligible correction 
by it. Excitations of stellar oscillations of the inspiraling stars,
or the 'dynamical tide', is another issue to
be considered. Reisenegger \& Goldreich (1994) and Lai (1994)
investigated resonant excitations of g-modes and their effects on 
the inspiral orbit. For slowly rotating stars g-modes frequencies 
may fall in the range as low as the orbital resonant frequency.
According to their results, however,  g-modes affect the orbit
negligibly, since the coupling between tidal potential and g-mode
eigenfunctions is rather weak.

As we have seen, counter-rotating f-modes of neutron stars pass 
the neutral points viewed from the asymptotic inertial frame 
if the star rotates sufficiently rapidly. Thus it may
suffer resonant excitations during the inspiral phase.\footnote{As seen 
in Bildsten \& Cutler, the tidal synchronization
of inspiraling components is unexpected. So the initial angular velocities
of both components are preserved during the inspiral. We here consider
the case in which at least one neutron star in the system has a sufficient
angular velocity to suffer the orbital resonance. Whether such systems
do exist actually or not is beyond our discussion here.}
Once the resonance condition is fulfilled, the f-mode is likely to be
a much more dangerous obstacle to the construction of the 
theoretical template, since the f-mode eigenfunction couples more 
strongly with the tidal potential. We here adopt the simple oscillator 
model by Reisenegger \& Goldreich (1994) and examine the effect of the 
mode excitation.

A sinusoidal external force operating on a star with a mass $M_1$
whose frequency fulfills the resonance condition, 
$\nu/m = n_{\mbox{\tiny orb}}$, where $n_{\mbox{\tiny orb}}$
is the orbital frequency, excites a mode whose energy amounts to,
\begin{equation}
	\varepsilon = \frac{(F\delta t)^2}{8 M_1},
\end{equation}
during the resonance time interval $\delta t$, which is typically 
the decay time of the orbit by gravitational radiation.
The external tidal force by the companion with a mass $M_2$ is estimated 
by the following formula:
\begin{equation}
	F = \frac{GM_1M_2}{R_1^2}\left(\frac{R_1}{a}\right)^{m+1} S,
\end{equation}
where $R_1$ is the stellar radius and $a$ is the separation of the 
binary system. The factor $S$ is the 'overlap integral' describing 
the efficiency of the tidal force on an eigenfunction defined by
\begin{equation}
	S = \int_{M_1} \sqrt{-g} u^t dr d\theta d\varphi 
\left(\frac{\epsilon}{c^2}\right) \vec\xi\cdot\vec P,
\end{equation}
where $g$ is the determinant of the metric of 
the background spacetime, $u^t$ the time component of the 4-velocity
of the unperturbed stellar fluid, $\vec\xi$ is the Lagrangian displacement and 
$\vec P = \nabla (r^mY_m^m(\theta,\varphi))$. 

The vibrational energy of an excited mode $\varepsilon$ is compared with
the orbital energy decrease by 
gravitational radiation $\Delta E$ in the time interval $\delta t$,
\begin{equation}
	\frac{\varepsilon}{\Delta E}
= \alpha_m[M_2/M_1;R_1]
\left(\frac{n_{\mbox{\tiny orb}}}{100\mbox{Hz}}\right)^{\frac{8m-23}{6}}
S^2,
\end{equation}
where the factor $\alpha_m$ depends on the mode number, the mass ratio of 
the components and the stellar radius. If this ratio is not negligible
compared with unity, the assumption that the binary orbit evolves 
solely by gravitational radiation from the orbital motion should be amended. 
If the stellar radius and the mass ratio of the components are 
assumed to be $R_1=10{\mbox{km}}$ and $M_2/M_1 = 1$, the factors
$\alpha_m$ are computed as follows:
\begin{equation}
\alpha_m = \cases{
	5\times 10^1 & (m=2) \cr
	0.4 & (m=3) \cr
	0.004 & (m=4) \cr
	3\times 10^{-5} & (m=5) \cr}
.
\end{equation}

If the mass ratio is larger, say $M_2/M_1 = 10$ (BH/NS case),
they are:
\begin{equation}
\alpha_m = \cases{
	2		& (m=2) \cr
	6\times 10^{-3} & (m=3) \cr
	2\times 10^{-5} & (m=4) \cr
	5\times 10^{-8} & (m=5) \cr}
.
\end{equation}

The overlap integral $S$ may be computed by using the eigenfunction
of the mode obtained by the Cowling approximation. 
We find that $S\simeq 1$ for the bar mode and $S$ is larger than
$0.6$, $0.3$, and $0.2$ for $m=3,4,$ and $5$ modes, respectively,
for all the models from the non-rotating state 
to the mass-shedding limit. \footnote{Here we made rough an estimation
which assumes the validity of Newtonian and slow-rotation-limit 
formulae used in the Reisenegger \& Goldreich in our investigation.
It is intended only to display that the overlap integral should be
nearly order of unity in the low order mode cases.}

\subsubsection{Neutron Stars with the Prograde Rotation to the 
Orbital Motion}
When the star under consideration rotates in the same direction as
its orbital motion, the resonance condition of 'counter-rotating' 
modes and the orbital motion may be fulfilled for the states 
beyond the neutral points of the modes.  In this case, the bar mode may 
not be significant except for the stars which rotate with velocities of 
nearly the mass-shedding limit.

As seen from Figs. 1 -- 3, for higher order modes, the orbital frequencies 
at the resonances, $n_{\mbox{\tiny orb}}=\nu/m$, may be under
$10^3\mbox{Hz}$ for the realistic EOS \footnote{Here we 
pay attention only to the models with the WFF3-NV EOS. }, 
even at the mass-shedding limit of the star.
Using the formula of $\varepsilon/\Delta E$ above, it is observed that
the higher modes than $m=4$ may not affect the orbital evolution
governed by gravitational radiation during the inspiral phase 
which will be observed by the gravitational wave antennas like LIGO/VIRGO. 
It is also seen that for systems with a large mass ratio like a BH/NS binary,
the energy deposited in the NS vibration is too small to affect
the binary orbit.

As for the $m=3$ mode, its excitation may affect the stability and
the evolution of the system if the star has a sufficiently large 
rotational frequency (for example, larger than $850$Hz for 
$M=1.0M_{\odot}$ ; larger than $1000$Hz for $M=1.4M_{\odot}$),
since the vibrational energy is no more negligible compared with 
the gravitational radiation energy loss from the system. Moreover
$n_{\mbox{\tiny orb}}$ at its resonance is in the range of sensitivity of 
LIGO/VIRGO detectors, the excitation may be observed as the discrepancy of 
the signal form from its theoretical template.

\subsubsection{Neutron Stars with the Retrograde Rotation to the
Orbital Motion}
If the star rotates in the retrograde direction against its orbital
motion, the resonance condition of the modes and the orbit requires 
the rotational frequency of the star to be lower than that of the 
neutral points. Thus the resonance condition can drop to the 
frequency range of LIGO/VIRGO sensitivity windows even for lower $m$ modes.

Significant is the fact that the bar mode which couples most strongly to 
the tidal potential can be resonant for a wide range of rotational 
frequency of the star. For $M=1.4M_{\odot}$ models, stars with 
rotational frequency above $400$Hz may suffer the resonant excitation 
on its inspiraling path in the gravitational wave antennas' sensitivity
window (for $M=1.0M_{\odot}$ case, this frequency may go down to $200$Hz). 
The factor $\alpha_2$ amounts to $50$ for an equal mass NS/NS binary, and
to $2$ even in BH/NS cases with the mass ratio $M_2/M_1=10$. 
Since the overlap integral $S\sim 1$ for the bar mode, the resonant 
excitation would affect the evolution of the binary orbit significantly.

Moreover the $m=3$ mode can be excited in the stars rotating with frequency
as low as $20$Hz (for $1.4M_{\odot}$ case). Thus theoretical templates of 
gravitational wave signals from compact binary systems containing NS 
with retrograde rotation should almost always take the resonant excitation 
of the counter-rotating f-modes into account.

\acknowledgments
SY would like to thank Prof. B. Schutz and Dr. C. Cutler for their warm 
and generous hospitality at Max-Planck-Institut f\"ur Gravitationsphysik 
(Albert-Einstein-Institut) in Potsdam where a part of the numerical computation
was done and this paper was prepared.
The authors are also grateful to the anonymous referee for useful suggestions.

\appendix
\section{Equilibrium Models with Lower Masses than the Mass-Shedding 
Limit Mass}
In Figures 14 and 15 we see a strange behavior of the 
rapidly rotating equilibrium models, i.e.,
the existence of equilibrium models in the right part of the figure
divided by the mass-shedding curve. It originates from the fact
that the rotational frequency becomes an improper index for
the stellar rotation for sufficiently rapid rotation models.
\begin{figure}[htbp]
\centering
\psfig{file=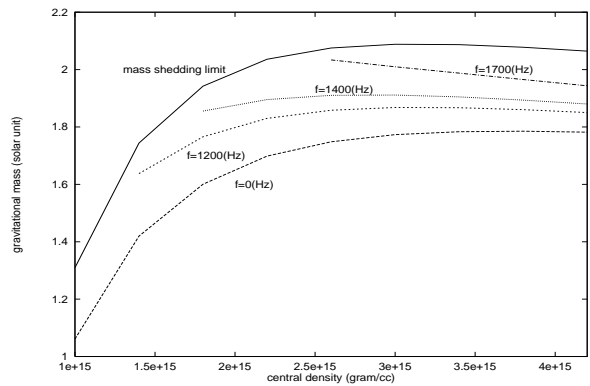,height=5cm,width=8cm,angle=-90}
\caption{The constant-rotational-frequency curves in the central
density $-$ gravitational mass plane. Here the models are constructed with 
the EOS Bethe-Johnson (neutron).\label{fig18}}
\end{figure}

\begin{figure}[htbp]
\centering
\psfig{file=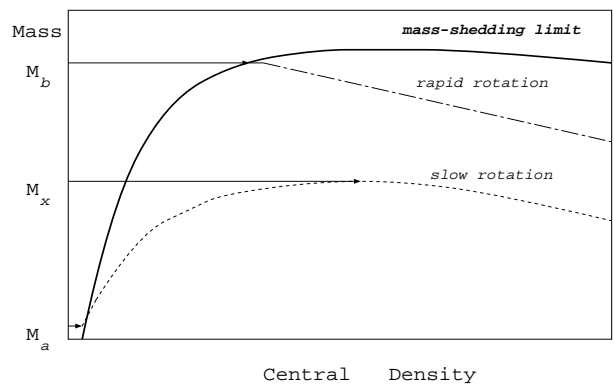,height=5cm,width=8cm}
\caption{Schematic figure of constant-rotational-frequency 
curves. The solid line is the mass-shedding curve. The dotted and 
dot-short-dashed lines respectively corresponds to the 
constant-rotational-frequency curves with slow and 
rapid rotation.  \label{fig19}}
\end{figure}
%
To see this we should know how gravitational mass changes with 
the central density of the star (Figure 18, Figure 19). In Figure 18
gravitational mass $M$ of the equilibrium models with the same 
angular frequency is shown as a function of the central energy 
density.
The solid line is the mass-shedding curve below which equilibrium 
models can be allowed to exist. 
Figure 19 is a schematic abstraction of the situation in Figure 18.

While the angular velocity is small the curve with a constant 
rotational frequency $f$ has a maximum in the allowed region, thus 
the mass of the mass-shedding configuration $M_a$ in Figure 19 
is {\it smaller} than the maximum mass $M_x$ for this value of 
angular frequency ($M_a\le M\le M_x$).
However, for the case with sufficiently rapidly rotation, 
the curve with constant angular frequency shows 
monotonically decreasing behavior with no local maximum in the allowed 
region. In this case, the gravitational mass at the
 mass-shedding model $M_b$ is {\it larger} than that of other 
equilibrium models with the same rotational frequency ($M_b \ge M$). 
This situation corresponds to the existence of the equilibrium model 
with smaller $M$ than the mass-shedding limit in Figures 14 and 15.


\begin{thebibliography}{DUM}
%
\bibitem[Andersson 1998]{NA98}
 Andersson, N. 1998, \apj, 502, 708
%
\bibitem[Andersson et al. 1998]{AKS98}
 Andersson, N., Kokkotas, K., \& Schutz, B. F. 1998, 
 \apj, Submitted.  ~(Preprint: astro-ph/9805225)
%
\bibitem[Arnett \& Bowers 1977]{AB77}
 Arnett, W. D., \& Bowers, R. L. 1977, \apjs, 33, 415
%
\bibitem[Baumgart \& Friedman 1986]{BF86}
 Baumgart, D., \& Friedman, J. L. 1986, Proc. R. Soc. Lond., A405, 65 
%
\bibitem[Bethe \& Johnson 1974]{BJ74}
 Bethe, H. A., \& Johnson, M. 1974, Nucl. Phys., A230, 1
%
\bibitem[Bildsten \& Cutler 1992]{BC92}
 Bildsten, L., \& Cutler, C. 1992, \apj, 400, 175
%
\bibitem[Chandrasekhar 1970]{CH70}
 Chandrasekhar, S. 1970, \prl, 24, 611
%
\bibitem[Comins 1979]{NC79}
 Comins, N. 1979, \mnras, 189, 255
%
\bibitem[Cutler et al. 1993]{CET93}
 Cutler, C. et al. 1993, \prl, 70, 2984
%
\bibitem[Friedman 1978]{JF78}
 Friedman, J. L. 1978, Comm. Math. Phys., 62, 247
%
\bibitem[Friedman \& Schutz 1978]{FS78}
 Friedman, J. L., \& Schutz, B. F. 1978, \apj, 222, 281
%
\bibitem[Komatsu et al. 1989]{KEH89}
 Komatsu, H., Eriguchi, Y., \& Hachisu, I. 1989, \mnras, 239, 153
%
\bibitem[Lai 1994]{DL94}
 Lai, D. 1994, \mnras, 270, 611
%
\bibitem[Lindblom 1986]{LL86}
 Lindblom, L. 1986, \apj, 303, 146
%
\bibitem[Lindblom et al. 1998]{LOM98}
 Lindblom, L., Owen, B. J., \& Morsink, S. M. 1998, \prl, 80, 4843
%
\bibitem[Morsink et al. 1998]{MSB98}
 Morsink, S. M., Stergioulas, N., \& Blattnig, S. R. 1998, \apj, Submitted.
~( Preprint: gr-qc/9806008 )
%
\bibitem[Nozawa et al. 1998]{NEA98}
 Nozawa, T., Stergioulas, N., Gourgoulhon, E., \& Eriguchi, Y. 1998,
\aap, in press.
%
\bibitem[Negel \& Vautherin 1973]{NV73}
 Negele, J. W., \& Vautherin, D. 1973, Nucl. Phys., A207, 298
%
\bibitem[Reisenegger \& Goldreich 1994]{RG94}
 Reisenegger, A., \& Goldreich, P. 1994, \apj, 426, 688
%
\bibitem[Stergioulas \& Friedman 1997]{SF97}
 Stergioulas, N., \& Friedman, J. L. 1997, \apj, 492, 301
%
\bibitem[Thorne 1969]{KT69}
 Thorne, K. S. 1969, \apj, 158, 997
%
\bibitem[Thorne 1994]{KT94}
 Thorne, K. S. 1994, in Relativistic Cosmology, 
ed. M. Sasaki ~(Tokyo:Universal Academy Press), 67
%
\bibitem[Wiringa et al. 1988]{WFF88}
 Wiringa, R. B., Fiks, V., \& Fabrocini, A. 1988, Phys. Rev., C38, 1010
%
\bibitem[Yoshida 1997]{SY97}
 Yoshida, S. 1997, Ph.D. thesis, University of Tokyo
%
\bibitem[Yoshida \& Eriguchi 1997]{YE97}
 Yoshida, S., \&  Eriguchi, Y. 1997, \apj, 490, 779
%
\end{thebibliography}
\end{document}